\documentclass[10pt,journal]{IEEEtran}
\usepackage{verbatim}
\usepackage{amsmath,bm}
\usepackage{graphicx}
\usepackage{algorithm}
\usepackage{algorithmic}
\usepackage{booktabs}
\usepackage{multicol}
\usepackage{cite}
\usepackage{mathrsfs}
\usepackage{color}
\usepackage{amsfonts,amssymb}

\usepackage[colorlinks,linkcolor=black,anchorcolor=black,citecolor=black]{hyperref}
\ifCLASSINFOpdf

\else

\fi

\begin{document}
%

\title{Cooperative Positioning for Sparsely Distributed High-Mobility Wireless Networks with EKF Based Spatio-Temporal Data Fusion}

\author{Yue~Cao, Shaoshi~Yang,~\IEEEmembership{Senior Member,~IEEE}, Xiao~Ma,
        and~Zhiyong~Feng,~\IEEEmembership{Senior Member,~IEEE}
\thanks{This work was supported in part by the Beijing Municipal
Natural Science Foundation under Grant L202012 and Grant Z220004, and in part by the Fundamental Research Funds for the Central
Universities under Grant 2020RC05. \textit{(Corresponding author: Shaoshi Yang.)}}
\thanks{Y. Cao, S. Yang and Z. Feng are with the School of Information and Communication Engineering, Beijing University of Posts and Telecommunications, and the Key Laboratory of Universal Wireless Communications, Ministry of Education,  Beijing 100876, 
China (e-mail: \{caoyue, shaoshi.yang, fengzy\}@bupt.edu.cn).}
\thanks{X. Ma is with China Academy of Launch Vehicle Technology, Beijing 100076, China (e-mail: xma\_casc@163.com).}

}

\markboth{Accepted by IEEE Communications Letters, Jun.~2023}%
{Shell \MakeLowercase{\textit{et al.}}: Bare Demo of IEEEtran.cls for Computer Society Journals}

\maketitle

\begin{abstract}
We propose a distributed cooperative positioning algorithm using the extended Kalman filter (EKF) based spatio-temporal data fusion (STDF) for a wireless network composed of sparsely distributed high-mobility nodes. Our algorithm first makes a coarse estimation of the position and mobility state of the nodes by using the prediction step of EKF. Then it utilizes the coarse estimate as the prior of STDF that relies on factor graph (FG), thus facilitates inferring \textit{a posteriori} distributions of the agents’ positions in a distributed manner. We approximate the nonlinear terms of the messages passed on the associated FG with high precision by exploiting the second-order Taylor polynomial and obtain closed-form representations of each message in the data fusion step, where temporal measurements by imperfect hardware are considered additionally. In the third stage, refinement of position estimate is performed by invoking the update step of EKF. Simulation results and analysis show that our EKF-STDF has a lower computational complexity than the state-of-the-art EKF-based algorithms, while achieving an even superior positioning performance in harsh environment.
\end{abstract}

\begin{IEEEkeywords}
Cooperative positioning, extended Kalman filter (EKF), high mobility, sparsely distributed, wireless localization.
\end{IEEEkeywords}

\IEEEpeerreviewmaketitle

\section{Introduction}
\IEEEPARstart{T}{he} location information of radio devices plays a crucial role in many emerging applications relying on wireless networks \cite{Wymeersch2009}. In harsh environments where the global navigation satellite system (GNSS) is denied, cooperative positioning (CP) \cite{win2011network,lv2015space,li2020robust,xiong2022cooperative,cao2022geo,cao2022spatial} is capable of providing the essential location information by solving a parameter estimation problem, where the wireless links between adjacent radio nodes are used to exchange spatial and/or temporal information, such as ranging measurements, angle of arrival \cite{response1,response2}, and inertial measurements. In particular, a class of distributed CP algorithms based on the factor graph (FG) have attracted intense attention\cite{Wymeersch2009,win2011network,lv2015space,cao2022geo,cao2022spatial}, and they are actually specific adaptations of the sum-product algorithm (SPA) \cite{Wymeersch2009}. In these algorithms, FG enables calculating the marginal \textit{a posteriori} probability density functions (PDFs) more efficiently, and the use of FG is more suitable for distributed implementations.

In some practical wireless networks, the nodes whose positions are unknown move fast and are sparsely distributed. However, the SPA-based algorithms ignore the mobility state of the nodes in the modeling process, hence resulting in increased positioning error in the high-mobility scenario over time. Additionally, in wireless networks comprising sparsely distributed nodes, the number of ranging measurements for agents (i.e., the nodes to be localized) may be insufficient, which inevitably leads to degraded positioning accuracy. To achieve more accurate location estimation in mobile networks, Huang \textit{et al.} proposed a state-transition and observability constrained extended Kalman filter (STOC-EKF) scheme \cite{kevin2015state}, which employed EKF to characterize the velocities of agents. However, it approximates the nonlinear system model as a linear model around selected linearization points, which inevitably degrades the positioning accuracy. In \cite{fan2018dynamic}, the authors proposed an SPA-aided CP scheme, dubbed SPA-EKF, which utilized EKF to estimate the velocities of agents as the \textit{a priori} information. However, it requires a large number of samples to approximate the nonlinear terms in the ranging measurements, hence imposing a high computational complexity. The authors of \cite{liang2022neural} proposed a graph neural network (GNN) enhanced belief propagation (BP) scheme for network navigation. \textcolor{black}{This scheme} refines the original messages propagated on FG by information learned from data driven GNNs. However, it also suffers from high complexity caused by massive particles and iterative message computations. Furthermore, all the approaches of \cite{kevin2015state,fan2018dynamic,liang2022neural} ignored the temporal-domain internal ranging measurements of agents, i.e., the distance traveled by each agent during a given period. 

Against the above background, we propose an EKF-based spatio-temporal data fusion (STDF) algorithm for solving the distributed CP problem in wireless networks composed of sparsely distributed high-mobility nodes. This problem is challenging and important for both commercial and military applications. Our contributions are summarized as follows. i) We develop a second-order Taylor polynomial (TP) based parametric method to approximate the nonlinear terms of both spatial and temporal messages passed on the FG.  As a beneficial result, closed-form representations for each type of messages are derived, and our method enjoys a competitive representation accuracy and a significantly reduced computational complexity than particle-based approaches. ii) We develop a joint prediction and refinement framework based on the integration of EKF and the second-order TP based parametric STDF to estimate the mobility state information of nodes, which is beneficial not only for compensating the lack of spatial ranging measurements caused by sparse distribution of nodes, but also for improving the accuracy of the prior knowledge used by STDF at each time slot. Thus, the proposed EKF-STDF alleviates performance degradation caused by the EKF-based model linearization and is more suitable to distributed CP than the data fusion module of \cite{fan2018dynamic} that relies on the original FG-free BP method.  iii) Simulation results demonstrate that the proposed EKF-STDF achieves a significantly higher positioning accuracy than the schemes of both \cite{kevin2015state} and \cite{fan2018dynamic} at an even lower computational complexity.

\section{System Model and Problem Formulation}
We consider a mobile network comprising $N$ agents and $A$ anchors in a GNSS-denied environment, and the transmission time is slotted. We denote the set of anchors and the set of particular agents from which agent \(i\) receives signals at time slot \(t\) by \(\mathbb{A}_i^{t}\) and \(\mathbb{U}_i^{t}\), respectively. In the two-dimensional (2D) scenario\footnote{Our algorithm can be extended to the three-dimensional space, but for convenience of presentation, in this paper we only discuss the 2D case.}, the state vector of agent \(i\) can be described as $\bm{s}_i^t=\left[\left(\bm{x}_i^t\right)^\text{T},\left(\bm{v}_{i}^t\right)^\text{T}\right]^\text{T}$, where \(\left(\cdot \right)^\text{T}\) denotes the transpose operation, \(\bm{x}_{i}^{t} \triangleq [x_{i}^{t}, y_{i}^{t}]^{\text{T}}\) and \(\bm{v}_{i}^{t} \triangleq [v_{i,x}^{t}, v_{i,y}^{t}]^{\text{T}}\) represent the position and the velocity of agent \(i\), respectively.

The noise-contaminated external ranging measurements obtained by agent \(i\) from node \(j\)  at time slot \(t\) is written as\footnote{The scenario considered in this paper is different from that of our previous work \cite{cao2022geo}, where we studied CP for wireless networks composed of static or slowly moving agents that operate in three-dimensional non-line-of-sight (NLOS)/LOS mixed environments.}
\begin{equation}
z_{j \rightarrow i}^{t}=d_{ij}^{t}+e^t_{j \rightarrow i},
\end{equation}
where \(d_{ij}^{t}\) is the Euclidean distance between agent \(i\) and node \(j\) at time slot \(t\), \(e^t_{j \rightarrow i}\sim\mathcal{N}\left(0, (\sigma_{j \rightarrow i}^t)^{2}\right)\) represents the Gaussian noise with zero-mean and variance \((\sigma_{j \rightarrow i}^t)^{2}\), and \(j \in \mathbb{A}_i^{t} \cup \mathbb{U}_i^{t}\). In addition, we assume the error of internal measurement $z^t_{i, \text{int}}$ obeys the Gaussian distribution with zero-mean and variance $(\sigma_{i, \textrm{int}}^t)^2$. We denote all the noisy ranging measurements (both external and internal) obtained by agent \(i\) at time slot \(t\) as \(\bm{z}_{i}^{t}\). Our goal is to estimate the positions of agent \(i\) given only these noisy measurements, i.e. \(p\left(\bm{x}_{i}^{t} \mid \bm{z}_{i}^{t}\right)\).
\section{The Proposed EKF-STDF Algorithm}
The proposed EKF-STDF comprises three stages: 1) prediction, 2) spatio-temporal data fusion, and 3) refinement.  
\subsection{Stage 1: Prediction}
The prediction stage is based on the prediction step of EKF, with the purpose of producing a coarse prediction concerning the state of the agent at the current time slot.

Consider a state transition model of EKF for agent \(i\) as\footnote{Our EKF-STDF is not limited to any specific state transition model.}
\begin{equation}
    \bm{s}_{i}^{t}=\bm{F} \bm{s}_{i}^{t-1}+\bm{w}_{i}^{t},
\end{equation}
where \(\bm{F}\) denotes the state transition matrix, satisfying
\begin{equation}
\bm{F}=\left[\begin{array}{cc}\bm{I}_{2} & \Delta T \bm{I}_{2} \\ \bm{0}_{2} & \bm{I}_{2}\end{array}\right],
\end{equation}
where \(\bm{I}_{2}\) and \(\bm{0}_{2}\) represent the identity matrix and the zero matrix of dimension 2, respectively; \(\Delta T\) is the duration of a single time slot; and \(\bm{w}_{i}^{t}\) represents the state transition noise that is modeled by the Gaussian vector with zero mean and covariance matrix \(\bm{Q}_{i}^{t}\). Then, the predictions about the mean and the covariance of the state $\bm{s}_{i}^{t}$, relying on the \textit{a posteriori} estimate at time slot $t-1$, are given by 
\begin{equation}
\hat{\bm{s}}_{i,\textrm{mean}}^{t|t-1}=\bm{F} \hat{\bm{s}}_{i,\textrm{mean}}^{t-1|t-1}, \label{al-1}
\end{equation}
\begin{equation}
\hat{\bm{P}}_{i}^{t|t-1}=\bm{F} \hat{\bm{P}}_{i}^{t-1|t-1} \bm{F}^{\mathrm{T}}+\bm{Q}_{i}^{t}.\label{al-2}
\end{equation}
Here, given observations up to and including at time $t-1$, $(\cdot)^{t-1|t-1}$  represents the \textit{a posteriori} estimate at time $t-1$, and $(\cdot)^{t|t-1}$ represents the \textit{a priori} estimate at time $t$, since $\hat{\bm{s}}_{i,\textrm{mean}}^{t|t-1}$ and $\hat{\bm{P}}_{i}^{t|t-1}$ are utilized by the data fusion stage as the \textit{a priori} distribution.
\subsection{Stage 2: Spatio-temporal data fusion}
We first factorize the \textcolor{black}{\textit{a posteriori}} distribution of the position concerning agent \(i\) at time slot \(t\) as
\begin{equation}
\begin{aligned} 
p\left(\bm{x}_{i}^{t} \mid \bm{z}_{i}^{t}\right) \ \propto \ & p\left(\bm{x}_{i}^{t}\right)   p\left(z_{i,\text{int}}^{t} | \bm{x}_{i}^{t}, \bm{x}_{i}^{t-1}\right)  \\ & \prod_{j \in \mathbb{A}_i^{t} \cup \mathbb{U}_i^{t}} p\left(z_{j \rightarrow i}^{t} | \bm{x}_{i}^{t}, \bm{x}_{j}^{t}\right).
\end{aligned}
\end{equation}
The \textit{a priori} distribution of $\bm{x}_i^t$, i.e., $p(\bm{x}_i^t)$, is obtained in Stage 1 upon assuming $\bm{x}_i^t \sim\mathcal{N}(\hat{\bm{x}}_{i}^{t},\hat{\bm{R}}_{i}^{t})$, where \(\hat{\bm{x}}_{i}^{t}=[\hat{x}_{i}^{t}, \hat{y}_{i}^{t}]\) denotes the position components\footnote{The velocity components are not involved in the data fusion.} in $ \hat{\bm{s}}_{i,\textrm{mean}}^{t|t-1}$, $\hat{\bm{R}}_{i}^{t}=\text{diag}\left((\hat{\sigma}_{i,x}^{t})^{2}, (\hat{\sigma}_{i,y}^{t})^{2}\right)$, while $(\hat{\sigma}_{i,x}^{t})^2$ and $(\hat{\sigma}_{i,y}^{t})^2$ are the $(1,1)$th and $(2,2)$th elements of \(\hat{\bm{P}}_{i}^{t|t-1}\), respectively. Then an iterative SPA is run on the FG of \(p\left(\bm{x}_{i}^{t} \mid \bm{z}_{i}^{t}\right)\), as depicted in Fig.~1\footnote{For more details about how to create and represent an FG, see \cite{cao2022spatial}.}. Since the proposed EKF-STDF is fully distributed, \textcolor{black}{let us consider the belief} about the $x$-component of the position concerning agent $i$ at time slot $t$ and iteration $l$, i.e., \textcolor{black}{${b}_{l}(x_{i}^{t})$}, as an example. It satisfies
\begin{equation}
	b_{l}(x_{i}^{t}) \propto \mu_{f_i^t\rightarrow x_i^t} \mu_{f_{i,\text{int}}^{t} \rightarrow x_{i}^{t}} \prod_{j \in \mathbb{U}_i^{t} \cup \mathbb{A}_i^{t}} \mu_{l, \phi_{j \rightarrow i} \rightarrow x_{i}^{t}}, \label{belief}
\end{equation}where we have $\mu_{f_i^t\rightarrow x_i^t}=p(x_i^t)\propto \mathcal{N}\left(\hat{x}_{i}^{t},(\hat{\sigma}_{i,x}^{t})^{2}\right)$, $\mu_{f_{i,\text{int}}^{t} \rightarrow x_{i}^{t}}$ denotes the temporal message obtained by internal measurements, and \textcolor{black}{$\mu_{l, \phi_{j \rightarrow i} \rightarrow x_{i}^{t}}$} represents the spatial messages passed from factor $\phi_{j \rightarrow i}$ to variable $x_{i}^{t}$ \textcolor{black}{at iteration $l$}, satisfying

\begin{equation}
\mu_{l, \phi_{j \rightarrow i} \rightarrow x_{i}^{t}}\propto\iiint \phi_{j \rightarrow i} b_{l}(x_{j}^{t}) b_{l}(y_{j}^{t})  d x_{j}^{t} d y_{j}^{t} d y_{i}^{t}. \label{m_spatial}
\end{equation}
In particular, when $j=k\in \mathbb{A}_i^{t} $, in \eqref{m_spatial} \textcolor{black}{we have $b_{l}(x_{k}^{t})=\delta\left(x_{k}^{t}-\mathrm{E}\{x_{k}^{t}\}\right)$, $b_{l}(y_{k}^{t})=\delta\left(y_{k}^{t}-\mathrm{E}\{y_{k}^{t}\}\right)$}, and the factor representing the likelihood function of $z^t_{k \rightarrow i}$, i.e., $\phi_{k \rightarrow i}$ satisfies
\begin{equation}
\phi_{k \rightarrow i}=\frac{1}{\sqrt{2 \pi \sigma_{k \rightarrow i}^{2}}} \exp \left\{-\frac{\left(z_{k \rightarrow i}^{t}-\left\|\bm{x}_{i}^{t}-\bm{x}_{k}^{t}\right\|_{2}\right)^{2}}{2 \sigma_{k \rightarrow i}^{2}}\right\}, \label{anchor-phi}
\end{equation}
where \(\delta(\cdot)\) is the Dirac delta function, $\mathrm{E}\{\cdot\}$ denotes the expectation, and \(\|\cdot\|_2\) represents the Euclidean norm.
\begin{figure}[t]
	\includegraphics[scale=0.3]{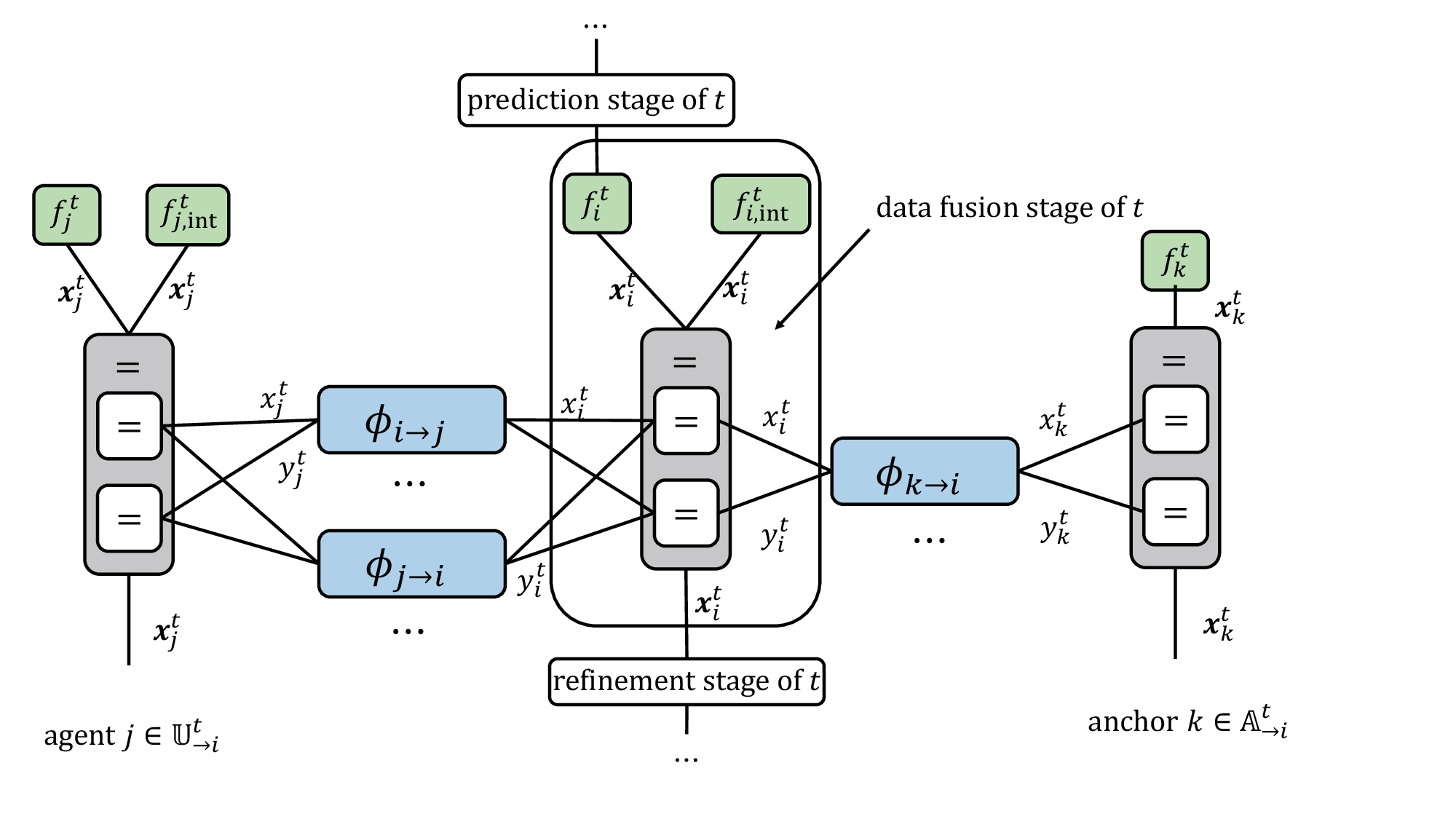}
	\centering
	\caption{FG of \(p\left(\bm{x}_{i}^{t} \mid \bm{z}_{i}^{t}\right)\), where agent \(j \in \mathbb{U}_i^{t}\), anchor \(k \in \mathbb{A}_i^{t}\), \(f_{i}^{t}=p\left(\bm{x}_i^t\right)\), \(f_{i,\text{int}}^t
    = p\left(z_{i, \text {int}}^{t} \mid \bm{x}_i^{t-1}, \bm{x}_i^{t}\right)\), and \(\phi_{j \rightarrow i}=p\left(z_{j \rightarrow i}^{t} \mid \bm{x}_{i}^{t}, \bm{x}_{j}^{t}\right)\).}
	\label{fig:1}       
\end{figure}
However, \eqref{m_spatial} involves integrals and it is difficult to obtain the closed-form expression due to \textcolor{black}{the nonlinear function of ${\bm x}_i^t$, i.e., $\left\|\bm{x}_{i}^{t}-\bm{x}_{k}^{t}\right\|_{2}$ of \eqref{anchor-phi}. To address this issue, $\left\|\bm{x}_{i}^{t}-\bm{x}_{k}^{t}\right\|_{2}$ is approximated by invoking the second-order TP, which is different from the method used in \cite{cao2022geo,cao2022spatial}. Upon substituting $b_{l}(x_{k}^{t})$, $b_{l}(y_{k}^{t})$ and \eqref{anchor-phi} into \eqref{m_spatial} we obtain\footnote{The positions of anchors are known constants during the iterations.}
\begin{equation}
\mu_{l, \phi_{k \rightarrow i} \rightarrow x_{i}^{t}}\propto\int \exp \left\{-\frac{\left(z_{k \rightarrow i}^{t}-\left\|\bm{x}_{i}^{t}-\mathrm{E}\{\bm{x}_{k}^{t}\}\right\|_{2}\right)^{2}}{2 \sigma_{k \rightarrow i}^{2}}\right\}  d y_{i}^{t}. \label{k2i}
\end{equation}
The second-order TP of $r_1 \triangleq \left\|\bm{x}_{i,l-1}^{t}-\mathrm{E}\{\bm{x}_{k}^{t}\}\right\|_2$ at $(\hat{x}_{i,l-1}^{t}, \hat{y}_{i,l-1}^{t})$ satisfies:
\begin{equation}
\begin{aligned}
r_1 \ & = r_1 (\hat{x}_{i,l-1}^{t}, \hat{y}_{i,l-1}^{t})+\sum_{\kappa \in \{x_i^t, y_i^t\}}\left(\kappa-\hat{\kappa}\right) \left(r_1\right)_{\kappa}^{\prime}(\hat{x}_{i,l-1}^{t}, \hat{y}_{i,l-1}^{t})
\\& + \frac{1}{2}\sum_{\kappa, \lambda \in \{x_i^t, y_i^t\}}(\kappa-\hat{\kappa}) (\lambda-\hat{\lambda}) \left(r_1\right)_{\kappa \lambda}^{\prime \prime}(\hat{x}_{i,l-1}^{t}, \hat{y}_{i,l-1}^{t}) + R_{r_1}, \label{r_1_second}
\end{aligned}
\end{equation}
where $R_{r_1}$ denotes the remainder term. Substituting \eqref{r_1_second} into \eqref{k2i}, the message passed from anchor $k$ to agent $i$ at iteration $l$ satisfies
\begin{equation}
\mu_{l, \phi_{k \rightarrow i} \rightarrow x_{i}^{t}} \propto \mathcal{N}(\frac{\beta_{k,l}}{2\alpha_{k,l}}, \frac{\gamma_{k,l}}{2\alpha_{k,l}}), \label{k2i_Gaussian}
\end{equation}
where we have
\begin{equation}
\left\{\begin{aligned}
\alpha_{k,l}= & 3\left(z^t_{k \rightarrow i} e_1^2-\|\boldsymbol{e}\|_2^3\right)\left(z^t_{k \rightarrow i} e_2^2-\|\boldsymbol{e}\|_2^3\right)-7 (z_{k \rightarrow i}^t)^2 e_1^2 e_2^2, \\
\beta_{k,l}= & 6\left(z^t_{k \rightarrow i} e_1 e_2 \hat{y}_{i,l-1}^{t}+z^t_{k \rightarrow i} e_1\|\boldsymbol{e}\|_2-z^t_{k \rightarrow i} e_2^2 \hat{x}_{i,l-1}^{t}\right. \\
& \left.+\mathrm{E}\left\{x_k^t\right\}\right) \left(\|\boldsymbol{e}\|_2^3-z^t_{k \rightarrow i} e_2^2\right)-14 z^t_{k \rightarrow i} e_1 e_2\left(\mathrm{E}\left\{y_k^t\right\} \right. \\
& \left.+z^t_{k \rightarrow i} e_2\|\boldsymbol{e}\|_2-z^t_{k \rightarrow i} e_1^2 \hat{y}_{i,l-1}^{t}+
z^t_{k \rightarrow i} e_1 e_2 \hat{x}_{i,l-1}^{t}\right), \\
\gamma_{k,l}= & 6\|\boldsymbol{e}\|_2^3 \sigma_{k \rightarrow i}^2\left(\|\boldsymbol{e}\|_2^3-z^t_{k \rightarrow i} e_1^2\right), \\
\boldsymbol{e}= & {\left[e_1, e_2\right]^{\text{T}}=\left[\hat{x}_{i,l-1}^{t}-\mathrm{E}\left\{x_k^t\right\}, \hat{y}_{i,l-1}^{t}-\mathrm{E}\left\{y_k^t\right\}\right]^{\text{T}}.}
\end{aligned}\right.
\end{equation}
When we have $j\in \mathbb{U}_i^{t}$, $\mu_{l, \phi_{j \rightarrow i} \rightarrow x_{i}^{t}}$ satisfies
\begin{equation}
\begin{aligned}
\mu_{l, \phi_{j \rightarrow i} \rightarrow x_{i}^{t}}  \propto & \iiint \exp \left\{-\frac{\left(z_{j \rightarrow i}^{t}-\left\|\bm{x}_{i}^{t}-\bm{x}_{j}^{t}\right\|_{2}\right)^{2}}{2 \sigma_{j \rightarrow i}^{2}}\right\}  
\\ & b_{l}(x_{j}^{t}) b_{l}(y_{j}^{t}) d x_j^t d y_j^t d y_i^t, \label{j2i}
\end{aligned}
\end{equation}}
where 
\begin{equation}
b_{l}(x_{j}^{t})=\frac{1}{\sqrt{2 \pi \sigma_{x_j^t}^2}} \exp \left\{-\frac{\left(x_j^t-\mathrm{E}\{x_j^t\}\right)^2}{2 \sigma_{x_j^t}^2}\right\}, \label{belief-xj}
\end{equation}
\begin{equation}
b_{l}(y_{j}^{t})=\frac{1}{\sqrt{2 \pi \sigma_{y_j^t}^2}} \exp \left\{-\frac{\left(y_j^t-\mathrm{E}\{y_j^t\}\right)^2}{2 \sigma_{y_j^t}^2}\right\}. \label{belief-yj}
\end{equation}
We utilize the second-order TP to approximate \(\left\|\bm{x}_{i}^{t}-\bm{x}_{j}^{t}\right\|_{2}\) at $(\hat{x}^t_{i,l-1}, \hat{y}^t_{i,l-1}, \hat{x}^t_{j,l-1}, \hat{y}^t_{j,l-1})$, and the message \textcolor{black}{\(\mu_{l, \phi_{j \rightarrow i} \rightarrow x_{i}^{t}}\) satisfies
\begin{equation}
\mu_{l, \phi_{j \rightarrow i} \rightarrow x_{i}^{t}} \propto \mathcal{N}(\frac{\beta_{j,l}}{2\alpha_{j,l}}, \frac{\gamma_{j,l}}{2\alpha_{j,l}}), \label{j2i_Gaussian}
\end{equation}}
where we have
\begin{equation}
\left\{\begin{array}{l}
\alpha_{j,l}=-q_2^2-\left(28 m_1 m_2^2 n_1 \sigma_{y_{j,l-1}^t}^4\right)^2,\\
\beta_{j,l}=28 m_1 m_2^2 n_1 \sigma_{y_{j,l-1}^t}^4\left(q_3-2\right)+q_2\left(q_3+2\right),\\
\gamma_{j,l}=18 m_1 n_1 q_1(q_2-28 m_1 m_2^2 n_1 \sigma_{y_{j,l-1}^t}^4),\\
\bm{g}=\left[g_1, g_2\right]^{\text{T}}=\left[\hat{x}_{i,l-1}^t-\hat{x}_{j,l-1}^t, \hat{y}_{i,l-1}^t-\hat{y}_{j,l-1}^t\right]^{\text{T}}, \\
q_1=3 n_1 \sigma_{y_{j,l-1}^t}^2(3 n_1+4 m_1 \sigma_{y_{j,l-1}^t}^2), \\
q_2=3 m_2^2 q_1-9\|\bm{g}\|_2^2 m_1 q_1, \\
q_3=3 n_1 m_1 m_2 \mathrm{E}\left\{y_{j,l-1}^t\right\} \sigma_{j \rightarrow i}^2 \sigma_{y_{j,l-1}^t}^2\\
\qquad-4 m_1 m_2 n_2 \sigma_{j \rightarrow i}^2 \sigma_{y_{j,l-1}^t}^2-9 q_1 n_3,\\
m_1=z^t_{j \rightarrow i}  g_1^2-\|\bm{g}\|_2^3, \\
m_2=2 z^t_{j \rightarrow i}  g_1 g_2, \\
n_1=\sigma_{j \rightarrow i}^2 \|\bm{g}\|_2^3, \\
n_2=-z^t_{j \rightarrow i}  g_2\|\bm{g}\|_2^2, \\
n_3=z^t_{j \rightarrow i}  g_1\|\bm{g}\|_2^2-(z^t_{j \rightarrow i})^2  g_1\|\bm{g}\|_2.
\end{array}\right.
\end{equation}

Similar to the treatment of the spatial messages \textcolor{black}{originating} from $j\in \mathbb{U}_i^{t}$, the $x$-component temporal message of agent $i$ from time slot $t-1$ to $t$, i.e., $\mu_{f_{i,\text{int}}^{t} \rightarrow x_{i}^{t}}$, satisfies\footnote{\textcolor{black}{Note that the temporal messages do not participate in the spatial iterations.}}
\begin{equation}
\mu_{f_{i,\text{int}}^{t} \rightarrow x_{i}^{t}} \propto  \mathcal{N}(\frac{\beta_i}{2\alpha_i}, \frac{\gamma_i}{2\alpha_i}), \label{temporal_Gaussian}
\end{equation}
where we have
\begin{equation}
\left\{\begin{array}{l}
\alpha_i=-q_5^2-\left(28 m_3 m_4^2 n_4 \sigma_{y_i^{t-1}}^4\right)^2,\\
\beta_i=28 m_3 m_4^2 n_4 \sigma_{y_i^{t-1}}^4\left(q_6-2\right)+q_5\left(q_6+2\right),\\
\gamma_i=18 m_3 n_4 q_4(q_5-28 m_3 m_4^2 n_4 \sigma_{y_i^{t-1}}^4),\\
\bm{h}=\left[h_1, h_2\right]^{\text{T}}=\left[\hat{x}_i^t-\hat{x}_i^{t-1}, \hat{y}_i^t-\hat{y}_i^{t-1}\right]^{\text{T}}, \\
q_4=3 n_4 \sigma_{y_i^{t-1}}^2(3 n_4+4 m_3 \sigma_{y_i^{t-1}}^2), \\
q_5=3 m_4^2 q_4-9\|\bm{h}\|_2^2 m_3 q_4, \\
q_6=3 n_4 m_3 m_4 \mathrm{E}\left\{y_i^{t-1}\right\}\left(\sigma_{i, \text {int}}^t\right)^2 \sigma_{y_i^{t-1}}^2\\
\qquad-4 m_3 m_4 n_5\left(\sigma_{i, \mathrm{int}}^t\right)^2 \sigma_{y_i^{t-1}}^2-9 q_4 n_6,\\
m_3=z_{i, \text {int}}^t  h_1^2-\|\bm{h}\|_2^3, \\
m_4=2 z_{i, \text {int}}^t  h_1 h_2, \\
n_4=\left(\sigma_{i, \text {int}}^t\right)^2 \|\bm{h}\|_2^3, \\
n_5=-z_{i, \text {int}}^t  h_2\|\bm{h}\|_2^2, \\
n_6=z_{i, \text {int}}^t  h_1\|\bm{h}\|_2^2-\left(z_{i, \text {int}}^t\right)^2  h_1\|\bm{h}\|_2.
\end{array}\right.
\end{equation}

Upon substituting \eqref{k2i_Gaussian}, \eqref{j2i_Gaussian}, \eqref{temporal_Gaussian} and the expression of \textcolor{black}{$p(x_i^t)$} into \eqref{belief}, and completing the iterations, we obtain
\begin{equation}
p(x_{i}^{t}|\bm{z}_{i}^{t})={b}_{l_{\text{max}}}(x_{i}^{t}) \propto \mathcal{N}(\mathrm{E}\{x_{i}^{t}|\bm{z}_{i}^{t}\}, \sigma_{x_{i}^{t}|\bm{z}_{i}^{t}}^{2}),
\end{equation}
where $l_{\text{max}}$ is the maximum number of iterations, and we have 
\begin{equation}
\begin{aligned}
\mathrm{E}\{x_{i}^{t}|\bm{z}_{i}^{t}\}&=\sigma_{x_{i}^{t} \mid \bm{z}^{t}}^{2}  \left[\frac{x_{i}^{t|t-1}}{\left(\sigma_{i,x}^{t|t-1}\right)^{2}}+ \sum_{k \in \mathbb{A}_i^{t}} \frac{\beta_{k,l_{\text{max}}}}{\gamma_{k,l_{\text{max}}}}\right.\\
&\left. \sum_{j \in \mathbb{U}_i^{t}} \frac{\beta_{j,l_{\text{max}}}}{\gamma_{j,l_{\text{max}}}}+\frac{\beta_i}{\gamma_i}\right],\label{24}
\end{aligned}
\end{equation}
\begin{equation}
\begin{aligned}
\sigma_{x_{i}^{t} \mid \bm{z}_{i}^{t}}^{2}&=\left[\frac{1}{\left(\sigma_{i,x}^{t|t-1}\right)^{2}}+ \sum_{k \in\mathbb{A}_i^{t}} \frac{2\alpha_{k,l_{\text{max}}}}{\gamma_{k,l_{\text{max}}}} \right.
\\ &\left. +\sum_{j \in\mathbb{U}_i^{t}} \frac{2\alpha_{j,l_{\text{max}}}}{\gamma_{j,l_{\text{max}}}}+\frac{2\alpha_i}{\gamma_i}\right]^{-1}.\label{25}
\end{aligned}
\end{equation}
The \(p(y_{i}^{t}|\bm{z}_{i}^{t})\) can be obtained in a similar manner. Therefore, the mean vector \(\bm{m}_{i}^{t}\) and covariance matrix \(\bm{R}_{i}^{t}\) concerning the position of agent \(i\) satisfy
\begin{equation}
\bm{m}_{i}^{t}=[\mathrm{E}\{x_{i}^{t}|\bm{z}_{i}^{t}\}, \mathrm{E}\{y_{i}^{t}|\bm{z}_{i}^{t}\}]^{\text{T}},\label{al-4}
\end{equation}
\begin{equation}
\bm{R}_{i}^{t}=\text{diag}\left(\sigma_{x_{i}^{t}|\bm{z}_{i}^{t}}^{2}, \sigma_{y_{i}^{t}|\bm{z}_{i}^{t}}^{2}\right).\label{al-5}
\end{equation}
\subsection{Stage 3: Refinement}
This step uses the update step of EKF to refine the \textit{a posteriori} distribution of agent \(i\) at time slot $t$. The measurement residual \(\Delta\bm{m}_{i}^{t}\) and its covariance matrix \(\bm{C}_{i}^{t}\) are given by
\begin{equation}
\Delta\bm{m}_{i}^{t}=\bm{m}_{i}^{t}-\bm{H}\hat{\bm{s}}_{i,\text{mean}}^{t|t-1},
\end{equation} and
\begin{equation}
\bm{C}_{i}^{t}=\bm{H} \hat{\bm{P}}_{i}^{t|t-1} \bm{H}^{\mathrm{T}}+\bm{R}_{i}^{t},
\end{equation}
\textcolor{black}{respectively,} where \(\bm{H}=\left[\begin{array}{cc}\bm{I}_{2} & \bm{0}_{2}\end{array}\right]\) is the observation matrix. Thus the near-optimal Kalman gain is given by
\begin{equation}
\bm{K}_{i}^{t}=\hat{\bm{P}}_{i}^{t|t-1} \bm{H}^{\mathrm{T}}\left(\bm{C}_{i}^{t}\right)^{-1},
\end{equation}
and it refines the marginal state distribution by weighting the measurement residual with
\begin{equation}
\hat{\bm{s}}_{i,\text{mean}}^{t|t}=\hat{\bm{s}}_{i,\text{mean}}^{t|t-1}+\bm{K}_{i}^{t} \Delta \bm{m}_{i}^{t}, \label{al-6}
\end{equation}
\begin{equation}
\hat{\bm{P}}_{i}^{t|t}=\hat{\bm{P}}_{i}^{t|t-1}-\bm{K}_{i}^{t} \bm{C}_{i}^{t}\left(\bm{K}_{i}^{t}\right)^{\mathrm{T}}.\label{al-7}
\end{equation}

To sum up, our EKF-STDF is presented in Algorithm \ref{alg1}.
\begin{algorithm} 
	\caption{The proposed EKF-STDF algorithm} 
	\label{alg1} 
	\begin{algorithmic}
	\footnotesize
		\REQUIRE The \textit{a priori} distribution \(\hat{\bm{s}}_{i, \text{mean}}^{0}, \hat{\bm{P}}_{i}^{0}, \forall i\).
		\ENSURE the refined marginal distribution \(\hat{\bm{s}}_{i, \text{mean}}^{t|t}, \hat{\bm{P}}_{i}^{t|t}, \forall i\).
        \FOR{agent $i$ = 1 to $N$}
        \STATE predict the \textit{a priori} distribution according to (\ref{al-1}) and (\ref{al-2}).
        \STATE compute the internal measurements based messages according to (\ref{temporal_Gaussian}).
		\FOR{iteration \(l\) = 1 to \(l_\text{max}\)}
		\STATE broadcast \(b_{l-1}(\bm{x}_{i}^{t})\).
        \STATE receive \(b_{l-1}(\bm{x}_{j}^{t})\), \(j \in \mathbb{A}_i^{t} \cup \mathbb{U}_i^{t}\) and compute the corresponding incoming messages according to \eqref{k2i_Gaussian} and \eqref{j2i_Gaussian}.
        \STATE using \eqref{24} and \eqref{25} to calculate the \textit{a posteriori} distribution.
        \ENDFOR
        \STATE obtain the position statistics \(\bm{m}_{i}^{t}\) and \(\bm{R}_{i}^{t}\) by (\ref{al-4}) and (\ref{al-5}).
		\STATE refine the \textit{a posteriori} state distribution using (\ref{al-6}) and (\ref{al-7}).
		\ENDFOR 
	\end{algorithmic} 
\end{algorithm}

\subsection{Computational Complexity}
\textcolor{black}{Since we consider a fully distributed network, it is sufficient to analyse the computational complexity imposed on a single agent during one time slot. Specifically, the proposed EKF-STDF has a complexity of \(\mathcal{O}\left(N_\text{rel}  l_{\text{max}}+\log_2 \left( N_\text{rel} l_{\text{max}}\right)\right)\), while each of the particle-based schemes, e.g., the SPA-EKF \cite{fan2018dynamic}, the particle-based SPAWN \cite{Wymeersch2009} and the NEBP \cite{liang2022neural}, has a complexity of \(\mathcal{O}\left(N_\text{rel}  N_\text{s}  l_{\text{max}}+\log_2 \left( N_\text{rel}  N_\text{s} l_{\text{max}} \right)\right)\). Here $N_\text{rel}$ is the number of neighbors of the agent considered, while $N_\text{s}$ denotes the number of particles required.}

\section{Simulation Results}
We evaluate the performance of our EKF-STDF against some representative positioning algorithms by numerical simulations in terms of root mean squared error (RMSE) between the estimated positions and the true positions. Specifically, consider a wireless network composed of 13 anchors in the area of $[0,3000]$ m $\times$ $[0,3000]$ m and a given number of agents (30 $\sim$ 60) uniformly placed in the area of $[100,2900]$ m $\times$ $[100,2900]$ m. The communication radius is set to 600~m. We assume that\footnote{Typically agents do not know in which direction they move, but they do know the distance they travel by internal ranging measurements.} the initial speed of each agent is 50 m/s and the random variation of the speed of each agent follows a Gaussian distribution with zero-mean and standard deviation of 5 m/s. We also assume that $\sigma_{j \rightarrow i}^2=0.01 d_{ij}^t$ and $\left(\sigma_{i, \text {int}}^t\right)^2=0.01\|\bm{x}_i^t-\bm{x}_i^{t-1}\|$. To ensure that the number of agents in the network remains constant, we place a new agent whenever an existing agent has left the considered area.

\begin{figure}[t]
	\centering\includegraphics[scale=0.4]{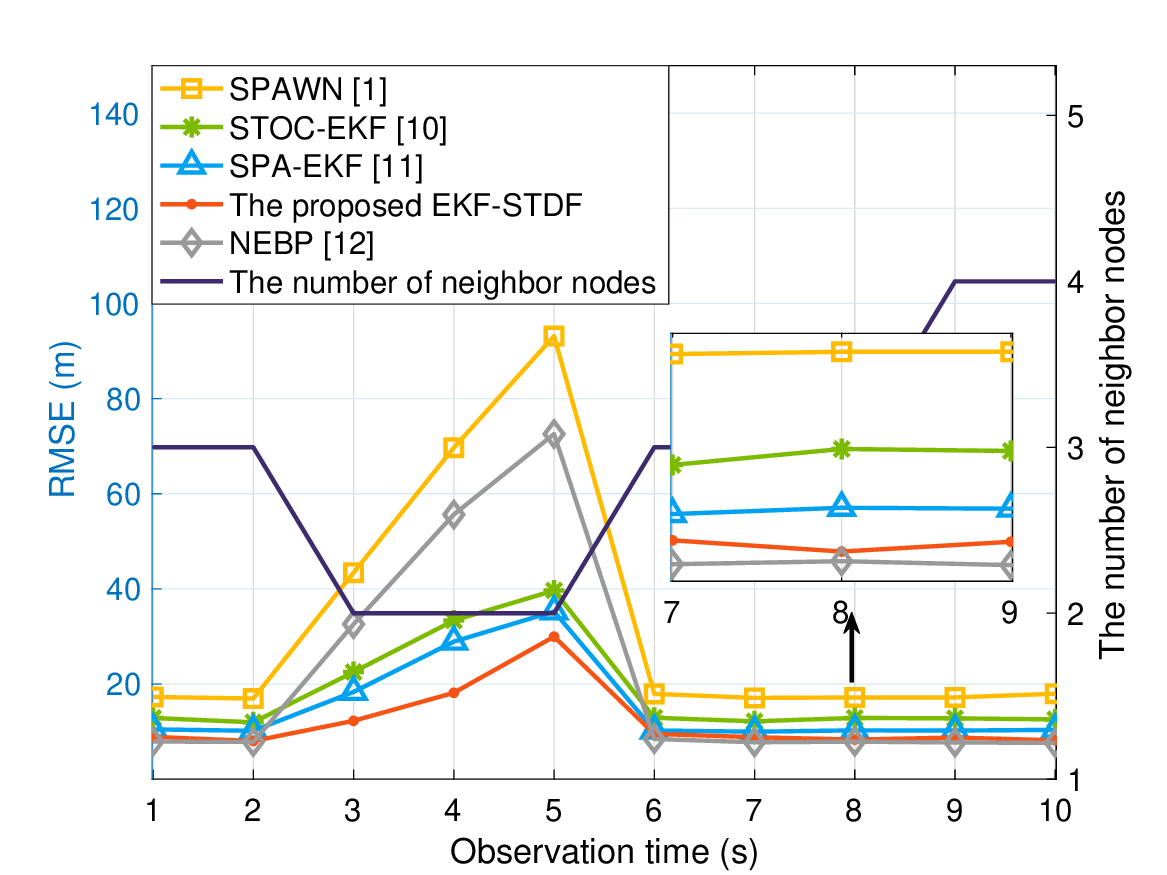}
	\caption{The positioning performance of a single agent under insufficient number of neighbors.}
	\label{fig:2}       
\end{figure}

In Fig.~\ref{fig:2} we compare the positioning performance of our EKF-STDF against the particle-based SPAWN\cite{Wymeersch2009}, STOC-EKF\cite{kevin2015state}, SPA-EKF\cite{fan2018dynamic} and NEBP\cite{liang2022neural}, by considering a single agent of interest that may have insufficient spatial ranging measurements from time to time. We set the number of agents to 40 and $l_\text{max}=30$. We have the following observations. Firstly, the performance of SPAWN and NEBP degrades rapidly when the agent does not have sufficient spatial ranging measurements, as characterized by the number of neighbor nodes, while the SPA-EKF, STOC-EKF, and EKF-STDF schemes are more robust to the deficiency of spatial ranging measurements. The results are compliant with our intuition that the prediction operation of EKF is helpful to improve the accuracy of the agent position estimation, and the high-precision prior values used in data fusion can reduce the ambiguity of the agent position estimation. Secondly, our EKF-STDF outperforms the SPA-EKF and STOC-EKF schemes whether the number of neighbor nodes is sufficient or not. This is attributed to the high-accuracy second-order TP approximation based STDF and the exploitation of the internal measurements based temporal information. Thirdly, the STOC-EKF is inferior to the SPA-EKF, because it linearizes the nonlinear observation model. This causes intrinsic performance degradation, especially in the high mobility scenario with insufficient neighbor nodes. Finally, when the number of neighbor nodes is sufficient, the positioning performance of NEBP is slightly superior to that of our EKF-STDF. However, the former is inapplicable to the scenario of insufficient neighbors.

\begin{figure}[t]
	\centering\includegraphics[scale=0.4]{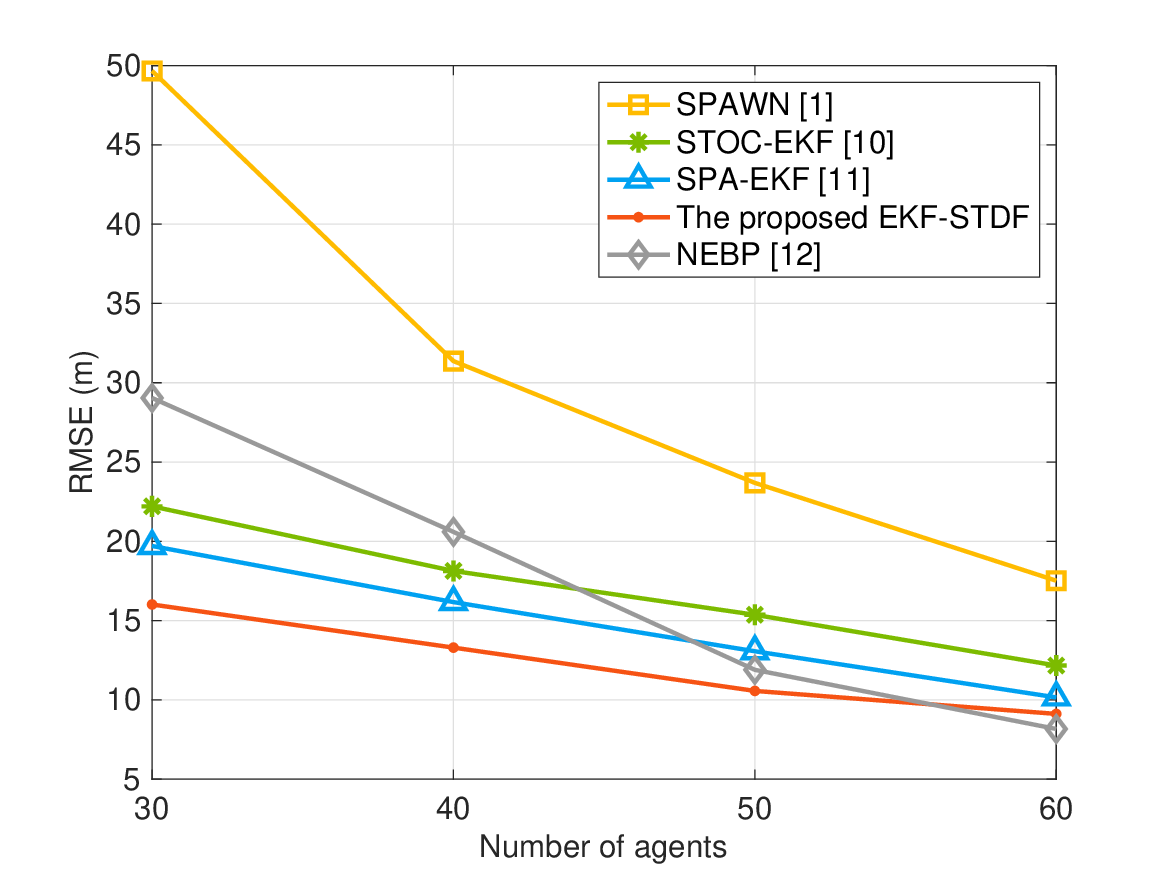}
	\caption{The positioning performance under different numbers of agents.}
	\label{fig:3}       
\end{figure}

Then we evaluate the positioning performance under different numbers of agents. We assume that the number of agents is increased from 30 to 60. Fig.~\ref{fig:3} shows how the number of agents influences the positioning performances of the above schemes. We have the following observations. Firstly, as the number of agents in the network increases, the gap between the EKF-based algorithms and SPAWN becomes smaller,  but the positioning performance of the latter remains inferior to that of the former. This observation is consistent with our intuition that when the \textcolor{black}{distribution of agents} is sparse, the lack of sufficient spatial ranging measurements leads to endogenous positioning bias in the regular SPA based SPAWN. \textcolor{black}{When} the number of \textcolor{black}{agents} in the network becomes larger, the existence of gap between SPAWN and EKF-based algorithms \textcolor{black}{is due to} the fact that the \textcolor{black}{prediction} and refinement \textcolor{black}{modules} of the latter make the \textit{a priori} and \textcolor{black}{the estimates} of agent positions more accurate. Secondly, our EKF-STDF outperforms NEBP when the number of agents \textcolor{black}{is} increased from 30 to 50, and the situation \textcolor{black}{is} reversed when the number of agents \textcolor{black}{reaches} 60. This indicates that when the number of agents is small, NEBP is limited by insufficient spatial ranging information, \textcolor{black}{which results in degraded} positioning accuracy. However, when the number of \textcolor{black}{agents} in the network is sufficient, the  \textcolor{black}{NEBP} outperforms the EKF-based schemes. Other observations are similar to \textcolor{black}{those obtained from} Fig.~\ref{fig:2}.

\section{Conclusion}
We have developed a low-complexity high-performance EKF-STDF algorithm to achieve a more attractive trade-off between the positioning accuracy and computational complexity for wireless networks that operate in sparsely distributed high-mobility environments. The proposed EKF-STDF exploits the prediction step of EKF to compute the \textit{a priori} state of the agents. Then aided by the \textit{a priori} position estimates and spatio-temporal ranging measurements, the data fusion stage infers the marginal distribution of the positions of the agents on the FG. In particular, we leverage the second-order TP to approximate the nonlinear functions in the messages passed on the FG in order to reduce the complexity. Finally, the refinement stage further enhances the positioning accuracy. Analysis and simulation results validated that our EKF-STDF has achieved competitive positioning performance with a lower computational complexity in the high mobility scenario with insufficient neighbor nodes.







\bibliographystyle{IEEEtran}
\bibliography{IEEEabrv,reference.bib}

\end{document}